\documentstyle[a4]{article}
\begin{document}

\noindent{\large\bf
4.2 Fundamentals of Electron-Photon Interaction 
in Nanostructures\footnote{
A slightly shortened version has been published as
section 4.2 of 
{\it Mesoscopic Physics and Electronics},
T. Ando et al.\ eds., Springer (1998),
pp. 140-155.
}
}

\begin{center}
Akira {\sc Shimizu}
\medskip

Institute of Physics, University of Tokyo \\
3-8-1 Komaba, Meguro-ku, Tokyo 153-8902, Japan

\end{center}

\renewcommand{\thefootnote}{\arabic{footnote}}
\renewcommand{\thesection}{\arabic{section}.}
\renewcommand{\thesubsection}{\arabic{section}.\arabic{subsection}}
\renewcommand{\thesubsubsection}{\arabic{section}.\arabic{subsection}.\arabic
{subsubsection}}

\renewcommand{\theequation}{\arabic{section}.\arabic{subsection}.\arabic
{equation}}

\renewcommand{\labelenumi}{(\alph{enumi})}

\setcounter{section}{4}
\setcounter{subsection}{2}


\def\k{{\bf k}}  \def\r{{\bf r}} \def\K{{\bf K}}  \def\R{{\bf R}}
\def\0{{\bf 0}} \def\rh{\mbox{\boldmath{$\rho$}}}
\def\A{{\bf A}} \def\e{\mbox{\boldmath{$\epsilon$}}}
\def\kp{{\bf k}_\parallel} 



We describe the fundamental equations for 
description 
of electron-photon interactions in insulating nanostructures. 
Since the theory for the nanostructures are based on the theory for
the bulk crystals, we shall describe 
elements of the theory for the bulk crystals 
\cite{knox}-\cite{b2} 
in the first three subsections.
It is important in discussions on optical phenomena 
that {\it both} the envelope function and the Wannier functions (or, 
equivalently,  
cell-periodic parts of the Bloch functions) should be 
taken into account.
The theory for the nanostructure is discussed in the fourth subsection.
In the final subsection, we briefly mention 
quantum optics in nanostructures.

\subsubsection{Electron and Hole Operators in Insulating Solids}
\label{EHO}

To describe quantum states in solids, it is customary to start from 
a mean-field approximation to electrons in a perfect crystal.
Then, according to the Bloch theorem, a single-body state of an electron in 
a solid crystal of volume $V$ takes the following form:
\begin{equation}
\phi_{b \k}(\r) = {1 \over \sqrt{V}} e^{i \k \cdot \r} u_{b \k}(\r),
\end{equation}
where $b$ labels bands (conduction band, heavy-hole band, and so on), and 
$u_{b \k}$ is the cell-periodic function which is periodic from cell to cell.
The spin index is omitted for the sake of simplicity.
The normalization is taken as
\begin{equation}
\int_V \! \! d^3 r \ 
\phi_{b \k}^*(\r) \phi_{b' \k'}(\r) 
= \delta_{b b'}\delta_{\k \k'},
\label{norm_bloch}\end{equation}
\begin{equation}
{1 \over v_0} \int_{v_0} \! \! d^3 \rho \ 
u_{b \k}^*(\rh) u_{b' \k}(\rh) 
= \delta_{b b'},
\label{norm_cell}\end{equation}
where $\rh$ denotes the coordinate vector in a unit cell, 
whose volume is ${v_0}$, and $\int \! \! d^3 \rho$ is over the unit cell.


We can construct the Wannier function 
which is localized around cell $\ell$ as
\begin{equation}
w_b (\r - \R_\ell)
\equiv
\sqrt{{v_0} \over V} {\sum_\k}' e^{-i \k \cdot \R_\ell} \phi_{b \k}(\r),
\end{equation}
where $\R_\ell$ is the center position of the cell, 
and ${\sum}'$ denotes the summation over the first Brillouin zone.
This function is normalized as
\begin{equation}
\int_V \! \! d^3 r \ 
w_b^* (\r - \R_\ell)
w_{b'} (\r - \R_{\ell'})
= \delta_{b b'}\delta_{\ell \ell'},
\label{norm_wannier}\end{equation}
and the inverse transformation is
\begin{equation}
\phi_{b \k}(\r) = 
\sqrt{{v_0} \over V} \sum_\ell e^{i \k \cdot \R_\ell} 
w_b (\r - \R_\ell).
\end{equation}
Note that both the Bloch functions and the Wannier functions form complete 
sets:
\begin{equation}
\sum_b {\sum_\k}' 
\phi_{b \k}^*(\r) \phi_{b \k}(\r')
= \delta(\r - \r'),
\end{equation}
\begin{equation}
\sum_b \sum_\ell w_b^* (\r - \R_\ell)
w_{b'} (\r' - \R_\ell)
= \delta(\r - \r').
\end{equation}
To describe many-body states, we introduce the electron field operator 
$\hat \psi(\r)$ which satisfies
\begin{equation}
\{ \hat \psi(\r), \hat \psi^\dagger(\r') \}
= \delta(\r - \r'),
\quad
\{ \hat \psi(\r), \hat \psi(\r') \}
=0.
\label{acr0}\end{equation}
We can expand $\hat \psi(\r)$ as
\begin{equation}
\hat \psi(\r) = 
\sum_b {\sum_\k}' 
\phi_{b \k}(\r) \hat c_{b \k}
=
\sum_b \sum_\ell 
w_b (\r - \R_\ell) \hat c_{b \ell}.
\end{equation}
We can easily show that $\hat c_{b \k}$ and $\hat c_{b \ell}$ satisfy 
\begin{equation}
\{ \hat c_{b \k}, \hat c_{b' \k'}^\dagger \} 
= \delta_{b b'} \delta_{\k \k'},
\quad
\{ \hat c_{b \k}, \hat c_{b' \k'} \}
=0, 
\end{equation}
\begin{equation}
\{ \hat c_{b \ell}, \hat c_{b' \ell'}^\dagger \} 
= \delta_{b b'} \delta_{\ell \ell'},
\quad
\{ \hat c_{b \ell}, \hat c_{b' \ell'} \} 
=0,
\end{equation}
and we can interpret 
 $\hat c_{b \k}^\dagger$ and $\hat c_{b \ell}^\dagger$ as the 
 creation operators of an electron whose wavefunction is $\phi_{b \k}(\r)$ 
and $w_b (\r - \R_\ell)$, respectively. 

We here consider a non-doped semiconductor or 
dielectric material. 
It is customary to define the creation operator of a hole by 
\begin{equation}
\hat h_{v \k \uparrow}^\dagger \equiv \hat c_{v -\k \downarrow}
\quad \mbox{when $v$ is a valence band},
\end{equation}
where only in this equation we recover the spin index for the reader's 
convenience.
The ground state is the vacuum of the holes
 and the conduction electrons:
\begin{equation}
\hat c_{c \k} |g \rangle = \hat h_{v \k} | g \rangle =0
\quad \mbox{for all $c$, $v$ and $\k$},
\label{gs}\end{equation}
where $c$ and $v$ stand for conduction and valence bands, respectively.
We hereafter call a conduction electron simply an electron.

It is convenient to decompose the fundamental field $\hat \psi$ into
the electron and hole parts, $\hat \psi_e$ and 
$\hat \psi_h$, respectively, as
\begin{equation}
\hat \psi(\r) =
\hat \psi_e(\r)+\hat \psi^\dagger_h(\r),
\end{equation}
\begin{equation}
\hat \psi_e(\r) \equiv \sum_c {\sum_\k}' \phi_{c \k}(\r) \hat c_{c \k},
\end{equation}
\begin{equation}
\hat \psi^\dagger_h(\r) 
\equiv \sum_v {\sum_\k}' \phi_{v \k}(\r) \hat c_{v \k}
= \sum_v {\sum_\k}' \phi_{v \k}(\r) \hat h^\dagger_{v -\k}.
\end{equation}
The anti-commutation relations are evaluated as
\begin{equation}
\{ \hat \psi_e(\r), \hat \psi_e^\dagger(\r') \}
= 
\sum_c {\sum_\k}' \phi_{c \k}(\r) \phi^*_{c \k}(\r') 
\equiv
\delta_e(\r - \r')
=
\delta^*_e(\r' - \r),
\end{equation}
\begin{equation}
\{ \hat \psi_h(\r), \hat \psi_h^\dagger(\r') \}
= 
\sum_v {\sum_\k}' \phi_{v \k}^*(\r) \phi_{v \k}(\r') 
\equiv
\delta_h(\r - \r')
=
\delta^*_h(\r' - \r),
\end{equation}
and other anti-commutators are zero.
Note that
\begin{equation}
\delta(\r - \r')=
\delta_e(\r - \r')+
\delta^*_h(\r - \r').
\end{equation}
Hence, $\delta_c$ and $\delta_v$ work as projection operators onto subspaces 
which are spanned by $\phi_{c \k}(\r)$'s and $\phi_{v \k}(\r)$'s, 
respectively.
{\it It is only in each subspace} 
that they work as delta functions 
and $\hat \psi^\dagger_e(\r)$ and $\hat \psi^\dagger_h(\r)$ can be 
interpreted as operators which create an electron and a hole, respectively, 
at the position $\r$.
One should always keep this limited meaning of $\delta_e$, $\delta_h$, 
$\hat \psi_e$ and $\hat \psi_h$ in mind to avoid inconsistencies.

\subsubsection{Effective-Mass Approximation}
\label{EMA}

A single-electron state can be obtained by operating 
$\hat c_{c \k}^\dagger$ on $| g \rangle$.
It is an eigenstate of  
the mean-field Hamiltonian $\hat {\cal H}_{mf}$ 
of the perfect crystal:
\begin{equation}
\hat {\cal H}_{mf} \hat c_{c \k}^\dagger | g \rangle
=
\varepsilon_c (\k) \hat c_{c \k}^\dagger | g \rangle,
\end{equation}
where we have taken $\hat {\cal H}_{mf} | g \rangle =0$, 
and $\varepsilon_c (\k)$ is called the band energy. 
A single-hole state, on the other hand, is obtained by operating 
$\hat h_{v \k}^\dagger$ on $| g \rangle$, and 
its energy is $-\varepsilon_v (\k)$:
\begin{equation}
\hat {\cal H}_{mf} \hat h_{v \k}^\dagger | g \rangle
=
-\varepsilon_v (\k) \hat h_{v \k}^\dagger | g \rangle,
\end{equation}

Suppose that $\varepsilon_c (\k)$ takes the minimum value at $\k = \k_{c}$, 
and $\varepsilon_v (\k)$ is maximum at $\k = \k_{v}$. 
The minimum energy cost for creating an electron in band $c$ and a hole in 
band $v$ is 
\begin{equation}
\varepsilon_c (\k_c) - \varepsilon_v (\k_v) \equiv E_{G cv},
\end{equation}
which is called the energy gap between bands $c$ and $v$.
In particular, the energy gap between the bottom (lowest) conduction band 
$c_m$ and the top (highest) valence band $v_m$ is simply called the energy 
gap, and is denoted by $E_G$. For simplicity, we hereafter assume that this 
gap is located at $\k=\0$, i.e., $\k_{c_m} = \k_{v_m} =\0$:
\begin{equation}
\varepsilon_{c_m} (\0) - \varepsilon_{v_m} (\0) \equiv E_{G}.
\end{equation}
We are interested in states with small $k$ in bands 
$c_m$ and $v_m$. We hereafter write $c_m$ and $v_m$ simply as 
$c$ and $v$.

When an additional single-body potential $V_1$, 
which may be due to impurities, external potential, and so on, 
are present the additional term appears in the Hamiltonian;
\begin{equation}
\hat {\cal H} = \hat {\cal H}_{mf} +
\int_V \! \! d^3 r \ \hat \psi^\dagger(\r) V_1(\r) \hat \psi(\r)
\equiv
\hat {\cal H}_{mf} +\hat {\cal H}_1.
\end{equation}
In this case, 
$\hat c_{c \k}^\dagger | g \rangle$'s 
and 
$\hat h_{v \k}^\dagger | g \rangle$'s 
 are no longer eigenstates, and
we must take their linear combinations to obtain true eigenstates.
The linear combination must be taken over different bands and different $\k$
's, in general.
However, the band mixing may be negligible if (i) the spatial variation of $V
_1$ is slow (i.e., the characteristic length of the variation $\gg$ lattice 
constants) and if 
(ii) the band of interest is well separated around $\k = \0$ 
from other bands. (This condition is usually 
satisfied for the conduction band. For the valence band, however,  
the degeneracy of heavy- and light-hole bands causes band mixing in 
many semiconductors. This point will be discussed later.)

In such a case, a single-electron eigenstate, 
\begin{equation}
\hat {\cal H} | e \eta \rangle =
\varepsilon_{e \eta} | e \eta \rangle,
\label{e-eigen}\end{equation}
which is
labeled by some set of quantum numbers $\eta$,
takes the following form:
\begin{equation}
| e \eta \rangle =
\int_V \! \! d^3 r \ \varphi_{c \eta}(\r) \hat \psi^\dagger(\r) | g \rangle
=
\int_V \! \! d^3 r \ \varphi_{c \eta}(\r) \hat \psi_e^\dagger(\r) | g \rangle
,
\end{equation}
\begin{equation}
\varphi_{c \eta}(\r)
=
{\sum_\k}' {\tilde F}_{c \eta}(\k) \phi_{c \k}(\r)
=
\sqrt{v_0} \sum_\ell F_{c \eta}(\R_\ell) w_c(\r - \R_\ell),
\label{ewf0}\end{equation}
where
\begin{equation}
F_{c \eta}(\r)
\equiv
{1 \over \sqrt{V}} {\sum_\k}' e^{i \k \cdot \r} {\tilde F}_{c \eta}(\k).
\end{equation}
The probability amplitude of finding an electron at the position $\r$ 
in the subspace which is spanned by $\phi_{c \k}(\r)$'s 
is evaluated as
\begin{equation}
\langle g | \hat \psi_e (\r) | e \eta \rangle =
\varphi_{c \eta}(\r).
\end{equation}
Therefore, we can interpret $\varphi_{e_\eta}(\r)$ as the single-body 
wavefunction of the electron in the subspace.
It is normalized as
\begin{equation}
1 = 
\langle e \eta | e \eta \rangle 
=
\int_V \! \! d^3 r \ |\varphi_{c \eta}(\r)|^2
=
{\sum_\k}' |{\tilde F}_{c \eta}(\k)|^2
=
v_0 \sum_\ell |F_{c \eta}(\R_\ell)|^2
=
\int_V \! \! d^3 r \ |F_{c \eta}(\r)|^2
\end{equation}
In a similar manner, 
a single-hole eigenstate takes the following form:
\begin{equation}
\hat {\cal H} | h \eta \rangle =
\varepsilon_{h \eta} | h \eta \rangle 
\label{h-eigen}\end{equation}
\begin{equation}
| h \eta \rangle 
=
\int_V \! \! d^3 r \ \varphi^*_{v \bar \eta}(\r) \hat \psi(\r) | g \rangle
=
\int_V \! \! d^3 r \ \varphi^*_{v \bar \eta}(\r) \hat \psi_h^\dagger(\r) 
| g \rangle,
\label{hwf0}\end{equation}
\begin{equation}
\varphi_{v  \eta}(\r)
=
{\sum_\k}' {\tilde F}_{v  \eta}(\k) \phi_{v \k}(\r)
=
\sqrt{v_0} \sum_\ell F_{v \eta}(\R_\ell) w_v(\r - \R_\ell),
\end{equation}
\begin{equation}
F_{v \eta}(\r)
\equiv
{1 \over \sqrt{V}} {\sum_\k}' e^{i \k \cdot \r} {\tilde F}_{v \eta}(\k),
\end{equation}
where 
$\bar \eta$ is the ``hole-conjugate" of $\eta$.
That is, we define $\bar \eta$ in such a way that it can be 
naturally interpreted as a set of 
quantum numbers of the hole: 
for a hydrogen-like state, for example, 
$\eta =(n,\ell,m,\sigma)$, where $(n,\ell,m)$ 
are the orbital quantum numbers and $\sigma$ denotes spin, 
we then take $\bar \eta =(n,\ell,-m,-\sigma)$. 
The probability amplitude of finding a hole 
at the position $\r$ 
in the subspace which is spanned by $\phi_{v \k}(\r)$'s is given by  
\begin{equation}
\langle g | \hat \psi_h (\r) | h \eta \rangle =
\varphi^*_{v \bar \eta}(\r).
\end{equation}
Therefore, we can interpret $\varphi^*_{v \bar \eta}(\r)$ as the 
single-body wavefunction of the hole which has 
the set of quantum numbers $\bar \eta$.
It is normalized as
\begin{equation}
1 = 
\langle h \eta | h \eta \rangle 
=
\int_V \! \! d^3 r \ |\varphi_{v \eta}(\r)|^2
=
{\sum_\k}' |{\tilde F}_{v \eta}(\k)|^2
=
v_0 \sum_\ell |F_{v \eta}(\R_\ell)|^2
=
\int_V \! \! d^3 r \ |F_{v \eta}(\r)|^2
\end{equation}

Under our assumptions (i) and (ii), 
$F_{b \eta} (\r)$ ($b = c, v$) becomes a slowly varying function of $\r$,  
and hence called 
the ``envelope function."
We note that if a function A(\r) is slowly varying and 
if a function $B(\r)$ 
is a cell-periodic function, then 
\begin{equation}
\int_V \! \! d^3 r \ A(\r) B(\r)
\simeq \sum_\ell A(\R_\ell) \int_{v_0} \! \! d^3 \rh \ B(\rh)
\simeq \left[ \int_V \! \! d^3 r A(\r) \right] \cdot 
\left[{1 \over v_0} \int_{v_0} \! \! d^3 \rh \ B(\rh) \right],
\label{formula0}\end{equation}
where $\int_{v_0} \! \! d^3 \rh$ means the integral over a unit cell.
Using this relation and the assumptions (i) and (ii),
we can derive the approximate eigenvalue equations for the envelope 
functions as 
\begin{equation}
\left[
\varepsilon_c \left( - i \nabla  \right)
+ V_1(\r)
\right]
F_{c \eta}(\r)
=
\varepsilon_{e \eta} 
F_{c \eta}(\r),
\end{equation}
\begin{equation}
\left[
- \varepsilon_v \left( - i \nabla \right)
- V_1(\r)
\right]
F_{v \eta}(\r)
=
\varepsilon_{h \bar \eta} 
F_{v \eta}(\r).
\end{equation}
In particular, when the bands have parabolic dispersions;
\begin{equation}
\varepsilon_c (\k) 
= \varepsilon_c (\0) + {\hbar^2 k^2 \over 2 m_e},
\label{para_e}\end{equation}
\begin{equation}
\varepsilon_v (\k) 
= \varepsilon_v (\0)-{\hbar^2 k^2 \over 2 m_h},
\label{para_h}\end{equation}
then we obtain the ``Schr\"odinger equations" for the envelope functions:
\begin{equation}
\left[
- {\hbar^2 \over 2 m_e} \nabla^2 + \varepsilon_c (\0)+ V_1(\r)
\right]
F_{c \eta}(\r)
=
\varepsilon_{c \eta} 
F_{c \eta}(\r),
\end{equation}
\begin{equation}
\left[
- {\hbar^2 \over 2 m_h} \nabla^2 
- \varepsilon_v (\0)- V_1(\r)
\right]
F_{v \eta}(\r)
=
\varepsilon_{h \bar \eta} 
F_{v \eta}(\r).
\end{equation} 
These eigenvalue equations are much easier to solve than 
the original Schr\"odinger equations (\ref{e-eigen}) and (\ref{h-eigen}),
because the complicated ``crystal potential" in ${\cal H}_{mf}$ are  
absorbed in the effective masses $m_e$, $m_h$ and the bottom and top energies
 of the bands, $\varepsilon_c (\0)$ and $\varepsilon_v (\0)$, respectively.
Once the envelope functions are thus obtained, the wavefunctions can be 
obtained from (\ref{ewf0}) or (\ref{hwf0}).
Because of this simplicity, 
the Schr\"odinger eqs.\ for the envelope functions are widely used in the 
studies of nanostructures.
However, care should be taken in discussions of optical properties, as we 
will explain in the subsequent subsections.

We next consider a state in which an electron-hole pair is excited.
In this case we must take account of the strong electron-hole interaction. 
If the band mixing caused by the interaction is negligible a pair state
(exciton state), 
which is labeled by  a set of 
quantum numbers $\eta$, 
can be written as
\begin{equation}
|x \eta \rangle 
=
\int_V \! \! d^3 r_e  \int_V \! \! d^3 r_h  \
\varphi_{x \eta}(\r_e, \r_h) 
\hat \psi_e^\dagger(\r_e)\hat \psi_h^\dagger(\r_h)
|g \rangle,
\end{equation}
\begin{eqnarray}
\varphi_{x \eta}(\r_e, \r_h)
&=&
{\sum_{\k_e}}' {\sum_{\k_h}}' 
{\tilde F}_{x \eta}(\k_e, \k_h) \phi_{c \k_e}(\r_e) \phi^*_{v \k_h}(\r_h)
\label{exwf_bloch}\\
&=&
v_0 \sum_j \sum_\ell 
F_{x \eta}(\R_j, \R_\ell)
w_c(\r_e - \R_j) w^*_v(\r_h - \R_\ell),
\label{exwf_wannier}\end{eqnarray}
where 
\begin{equation}
F_{x \eta}(\r_e, \r_h)
=
{1 \over V}
{\sum_{\k_e}}' {\sum_{\k_h}}' 
e^{i \k_e \cdot \r_e - i \k_h \cdot \r_h} 
{\tilde F}_{x \eta}(\k_e, \k_h).
\label{F_vs_f}\end{equation}
The 
probability amplitude of finding 
an electron at the position $\r_e$ and 
a hole at $\r_h$ is evaluated as
\begin{equation}
\langle g | \hat \psi_h (\r_h) \hat \psi_e (\r_e) 
| \varphi_{x \eta} \rangle 
=
\varphi_{x \eta}(\r_e, \r_h).
\end{equation}
Therefore, $\varphi_{x \eta}(\r_e, \r_h)$ can be 
interpreted as the wavefunction of the pair.
It is normalized as
\begin{eqnarray}
1 &=& 
\langle x \eta | x \eta \rangle 
=
\int_V \! \! d^3 r_e \int_V \! \! d^3 r_h \ 
|\varphi_{x \eta}(\r_e, \r_h)|^2
=
{\sum_{\k_e}}' {\sum_{\k_h}}' 
 |{\tilde F}_{x \eta}(\k_e, \k_h)|^2
\nonumber\\
&=&
v_0^2 \sum_j \sum_\ell |F_{x \eta}(\R_j, \R_\ell)|^2
=
\int_V \! \! d^3 r_e \int_V \! \! d^3 r_h \ 
|F_{x \eta}(\r_e, \r_h)|^2.
\label{ex_norm}\end{eqnarray}
Note that $\varphi_{x \eta}(\r_e, \r_h)$ is not required to be 
antisymmetric under  
interchange of $\r_e$ and $\r_h$.
That is, we can treat an electron and a hole 
as if they were different particles (as far as we can neglect the band 
mixing).

The forms (\ref{exwf_bloch})-(\ref{exwf_wannier}) of the wavefunction 
can describe both the Wannier exciton (for which the effective Bohr radius 
$a_B^*$ $\gg$ lattice constant $a$) and the Frenkel 
exciton (for which $a_B^* \sim a$).
We here consider the case of the Wannier exciton, and 
assume, for simplicity, the parabolic dispersions
(\ref{para_e})-(\ref{para_h}).
Then, in a perfect crystal, the total 
crystal momentum $\K$ of the pair becomes a good quantum number, 
and we write $\eta = (\K, \nu)$, where 
$\nu$ denotes the set of the other quantum numbers. 
Considering that $\K$ should be shared between the electron and hole 
in proportion to their masses, 
we can take ${\tilde F}_{x \eta}$ in the following form:
\begin{equation}
{\tilde F}_{x \eta} (\k_e, \k_h)
=
{\sum_\k}'
\tilde U_\nu (\k) 
\delta_{\k_e, \k+{m_e \over M} \K}
\delta_{\k_h, \k-{m_h \over M} \K}
\equiv
{\tilde F}_{\K \nu} (\k_e, \k_h),
\end{equation}
where $\tilde U_\nu$ is a function to be determined, 
and $M \equiv m_e + m_h$.
Inserting this form into Eq.\ (\ref{F_vs_f}), we obtain 
\begin{equation}
F_{x \eta}(\r_e, \r_h)
=
{1 \over \sqrt{V}}
\exp[i \K \cdot \R] \
U_{\nu}(\r_e - \r_h)
\equiv
F_{\K \nu}(\r_e, \r_h)
\label{exef1}\end{equation}
where we have introduced the center-of-mass coordinate,
\begin{equation}
\R \equiv {m_e \r_e + m_h \r_h \over M}, 
\end{equation}
and 
\begin{equation}
U_\nu (\r) \equiv 
{1 \over \sqrt{V}} {\sum_\k}' 
e^{i \k \cdot \r} \tilde U_\nu(\k).
\end{equation}
We can interpret Eq.\ (\ref{exef1}) as the product of 
the center-of-mass envelope function $e^{i \K \cdot \R}/\sqrt{V}$ and 
the relative-motion envelope function $U_\nu$.
From (\ref{ex_norm}) we see that each of them are normalized to 1:
\begin{equation}
1 
= \int_V \! \! d^3 R 
\left| 
{1 \over \sqrt{V}} \exp[i \K \cdot \R] 
\right|^2
= \int_V \! \! d^3 r |U_\nu (\r)|^2
= {\sum_\k}' |\tilde U_\nu (\k)|^2.
\end{equation}

The functional form of $U_\nu$ depends crucially on 
the electron-hole interaction.
The form of the e-h interaction was reviewed, for example, in \cite{knox}.
It was shown there that the form is nontrivial for 
$
r <\!\!\!\!\!\!_{_\sim} a_B^*
$ because of various exchange potentials and 
the finite size of the electron and hole.
Since we are considering Wannier excitons,
the long-range parts of the interactions are more important than the 
short-range parts.
The dominant e-h interaction would then be the screened 
Coulomb attraction $-e^2 / \epsilon r$, where 
$\epsilon$ is the dielectric constant.
When the e-h interaction is approximated by it, 
the ``Shr\"odinger equation" for $U_\nu$ takes the familiar form;
\begin{equation}
\left[
- {\hbar^2 \over 2 \mu} \nabla^2 - {e^2 \over \epsilon r}
\right]
U_\nu
=
E_\nu U_\nu, 
\label{eq_for_U}\end{equation}
where $\mu$ is the reduced mass,
$
1/\mu = 1/m_e + 1/m_h.
$
By solving the simple eigenvalue equation (\ref{eq_for_U}), we can calculate
the eigen-energy $E_{\K \nu}$ and the wavefunction $\varphi_{\K \nu}$ 
of the state $| \K \nu \rangle$.
That is, 
\begin{equation}
E_{\K \nu} = E_G + E_\nu + {\hbar^2 K^2 \over 2 M}.
\end{equation}
We see that the exciton energy consists of the 
band gap energy $E_G$, 
the energy of the e-h relative motion $E_\nu$, and 
the energy of the center-of-mass motion $\hbar^2 K^2 / 2 M$.
The binding energy of the e-h pair is $-E_\nu$. 
We can obtain the wavefunction $\varphi_{x \eta}(\r_e, \r_h)
\equiv
\varphi_{\K \nu}(\r_e, \r_h)$ by inserting 
the eigenfunction $U_\nu$ into 
Eq.\ (\ref{exef1}), and the resulting expression into (\ref{exwf_wannier}):
\begin{eqnarray}
\varphi_{\K \nu}(\r_e, \r_h)
=
v_0 \sum_j \sum_\ell 
{1 \over \sqrt{V}}
\exp \left[
i \K \cdot {m_e \R_j + m_h \R_\ell \over M}
\right]
U_{\nu}(\R_j - \R_\ell)
&&
\nonumber\\
\times \ 
w_c(\r_e - \R_j) w^*_v(\r_h - \R_\ell).
\qquad &&
\label{exwf1}
\end{eqnarray}
Or, we may insert $\tilde U_\nu$ into (\ref{exwf_bloch}) to get
\begin{equation}
\varphi_{\K \nu}(\r_e, \r_h)
=
{1 \over \sqrt{V}}
\exp[i \K \cdot \R] \
S_{\K \nu}(\r_e, \r_h),
\label{exwf2}
\end{equation}
\begin{equation}
S_{\K \nu}(\r_e, \r_h)
= 
{1 \over \sqrt{V}} {\sum_\k}'
\exp[i \k \cdot (\r_e - \r_h)] \
\tilde U_\nu(\k) \
u_{c \k+{m_e \over M}\K}(\r_e) 
u^*_{v \k-{m_h \over M}\K}(\r_h).
\label{S}\end{equation}
Note that we can 
replace $u_c(\r_e)$ and $u_v(\r_h)$ with 
$u_c(\rh_e)$ and $u_v(\rh_h)$, respectively,
where $\rh_e$ and $\rh_h$ denote the position vectors 
which are reduced in a unit cell (i.e., 
they are $\r_e$ and $\r_h$ measured from the center of  
the unit cells in which $\r_e$ and $\r_h$ are located, respectively).

Actually, except for high-quality crystals kept at a low-enough temperature,
the crystal momentum $\K$ of the center-of-mass motion is not conserved 
because the exciton is scattered by phonons, impurities and/or defects, and 
the exciton wavefunctions are modified.
For example, when the total potential of impurities is  
slowly-varying 
(i.e., 
characteristic length of the variation $\gg a_B^*$)  the wavefunction of 
an exciton state $| N \nu \rangle$
which is trapped by the impurities  may be obtained as superposition of 
$\varphi_{\K \nu}(\r_e, \r_h)$ over various $\K$'s,
\begin{eqnarray}
\varphi_{N \nu}(\r_e, \r_h)
&=&
{\sum_\K}' {\tilde G}_{N \nu}(\K) \varphi_{\K \nu}(\r_e, \r_h)
\label{exwf_trap}\\
&=&
v_0 \sum_j \sum_\ell 
G_{N \nu} \left( {m_e \R_j + m_h \R_\ell \over M} \right) 
U_{\nu}(\R_j - \R_\ell)
\nonumber \\
& & \qquad \times \ 
w_c(\r_e - \R_j) w^*_v(\r_h - \R_\ell),
\label{exwf_trap_wannier}\end{eqnarray}
where ${\tilde G}_{N \nu}$ is a superposition coefficient,
with $N$ being a quantum number, and 
\begin{equation}
G_{N \nu} (\R) 
\equiv
{1 \over \sqrt{V}}
{\sum_\K}' {\tilde G}_{N \nu}(\K) 
e^{i \K \cdot \R}
\end{equation}
can be regarded as 
the envelope function of the trapped center-of-mass motion.
It is normalized as
\begin{equation}
\int_V G_{N \nu}^*(\R) G_{N' \nu}(\R) d^3 R 
=
{\sum_\K}' {\tilde G}_{N \nu}^*(\K) {\tilde G}_{N' \nu}(\K)
=\delta_{N,N'}.
\label{norm_g}\end{equation}
The trapping 
may be viewed as the Anderson localization of the exciton, 
where the size $\ell_G$ of the spatial extension of $G_{N \nu} (\R)$ is the 
localization length $\ell_{loc}$.
We may also use the form (\ref{exwf_trap}) or (\ref{exwf_trap_wannier})
as an approximate 
wavefunction when the quantum coherence of the center-of-mass motion 
is broken by scatterings by phonons or other excitons. 
In this case $\ell_G$ is 
the phase breaking length $\ell_\phi$.
Both the localization and the phase breaking may be important 
in real systems, and $\ell_G$ would  be given roughly by
\begin{equation}
\ell_G \simeq \min [\ell_{loc}, \ \ell_\phi].
\label{ell_G}\end{equation}
It will be shown later that the oscillator strength of 
the exciton is proportional to $\ell_G^3$ when $\ell_G$ is 
less than the wavelength of the optical field.
%

In a similar manner, states with two or more e-h pairs 
can be written in 
terms of two- or more-particle wavefunctions as
\begin{eqnarray}
&&
\int_V \! \! d^3 r_e^1 \int_V \! \! d^3 r_e^2 \cdots
\int_V \! \! d^3 r_h^1 \int_V \! \! d^3 r_h^2 \cdots 
\
\varphi(\r_e^1, \r_2^2,\cdots, \r_h^1, \r_h^2, \cdots) 
\nonumber \\
&& \qquad \qquad \times \
\hat \psi_e^\dagger(\r_e^1) \hat \psi_e^\dagger(\r_e^2) \cdots
\hat \psi_h^\dagger(\r_h^1) \hat \psi_h^\dagger(\r_h^2) \cdots
|g \rangle.
\end{eqnarray}
General states are linear combinations and/or classical mixture 
of such states with zero, one, and 
more e-h pairs.

\subsubsection{Optical Matrix Elements}
\label{OME}

The envelope functions should not be taken as the true 
wavefunctions because optical phenomena are frequently accompanied with 
transitions between the cell-periodic parts $u_{b \k}$'s of the Bloch 
functions.
That is, optical matrix elements depend not only on 
the envelope functions but also on the 
functional forms of $u_{b \k}$'s (or, equivalently, 
of the Wannier functions).
The latter dependence becomes crucial when nonlinear optical processes or 
intraband transitions are discussed.
Unfortunately, this point was disregarded in much literature, 
and unphysical results were sometimes reported.
We here present typical matrix elements for the states described in the 
previous subsection.

The interaction with an external 
optical field $\A(\r, t)$ in the Coulomb gauge takes the following form:
\begin{eqnarray}
\hat {\cal H}_I
&=&
\int_V \! \! d^3 r \ 
\hat \psi^\dagger (\r)
\left[
- {\hbar e \over i m c} \A \cdot \nabla
+ {e^2 \over 2 m c^2} A^2
\right]
\hat \psi (\r),
\\
&=&
\hat {\cal H}_I^{(1)}+\hat {\cal H}_I^{(2)},
\label{H_I}
\end{eqnarray}
where $\hat {\cal H}_I^{(1)}$ is linear in $\A$, whereas 
$\hat {\cal H}_I^{(2)}$ is quadratic.
To evaluate various matrix elements of 
$\hat {\cal H}_I^{(1)}$ and $\hat {\cal H}_I^{(2)}$, 
it is useful to derive formulas 
for matrix elements of a general operator of the form
\begin{equation}
\hat Q
=
\int_V \! \! d^3 r \int_V \! \! d^3 r' \ 
Q(\r, \r')
\hat \psi^\dagger (\r)
\hat \psi (\r').
\end{equation}
In some cases, its expectation value in the ground state $| g \rangle$ 
does not vanish.
Such a non-zero expectation value in the ground state can be absorbed 
by redefinition of the operator. This means a 
change of the zero of energy when $\hat Q$ is a part of the Hamiltonian.
The redefinition is most conveniently accomplished by 
the ``normal order" with respect to $\hat \psi_e$ and $\hat \psi_h$. 
The normally-ordered $\hat Q$ is 
\begin{eqnarray}
: \hat Q :
&=&
\int_V \! \! d^3 r \int_V \! \! d^3 r' \ 
Q(\r, \r')
: \hat \psi^\dagger (\r) \hat \psi (\r') :
\nonumber\\
&=&
\int_V \! \! d^3 r \int_V \! \! d^3 r' \ 
Q(\r, \r')
\left[ 
\hat \psi_e^\dagger (\r) \hat \psi_e (\r')
+ \hat \psi_e^\dagger (\r) \hat \psi_h^\dagger (\r')
\right.
\nonumber\\
& & \qquad \qquad 
\left.
+ \hat \psi_h (\r) \hat \psi_e (\r')
- \hat \psi_h^\dagger (\r') \hat \psi_h (\r)
\right],
\end{eqnarray}
which differs from $\hat Q$ only by a c-number term.
Using Eq.\ (\ref{gs}) and the anti-commutation relations, 
we can easily show that
\begin{equation}
\langle g | : \hat Q : | g \rangle = 0,
\label{formula1}\end{equation}
\begin{equation}
\langle x \eta | : \hat Q : | g \rangle
=
\int_V \! \! d^3 r_1  \int_V \! \! d^3 r_2  \
\varphi^*_{x \eta}(\r_1, \r_2) 
Q(\r_1, \r_2),
\label{formula2}\end{equation}
\begin{eqnarray}
\langle x \eta | : \hat Q : | x \eta' \rangle
=
\int_V \! \! d^3 r_1  \int_V \! \! d^3 r_2 \int_V \! \! d^3 r_3 \
[
\varphi^*_{x \eta}(\r_1, \r_3) 
Q(\r_1, \r_2)
\varphi_{x \eta'}(\r_2, \r_3)
&&
\nonumber\\ 
-
\varphi^*_{x \eta}(\r_3, \r_1) 
Q(\r_2, \r_1)
\varphi_{x \eta'}(\r_3, \r_2)
].
\quad &&
\label{formula3}\end{eqnarray}
These formulas enable us to evaluate various matrix elements 
from the two-body wavefunction $\varphi_{x \eta}(\r_e, \r_h)$.

For an optical field whose photon energy is of the order of $E_G$ or less, 
the wavelength $\lambda$ of $\A$ is hundreds nano meters or longer.
When the spatial extension of $\varphi_{x \eta}(\r_e, \r_h)$ is 
smaller than $\lambda$ 
we can approximate  matrix elements of $: {\cal H}_I :$ as
\begin{equation}
\langle x \eta | 
: {\cal H}_I^{(1)} :
| g \rangle  
\simeq
- {\hbar e \over i m c} \A(\r_{x \eta})
\int_V \! \! d^3 r \ 
\langle x \eta | 
: 
\hat \psi^\dagger (\r)
\ \e \cdot \nabla \
\hat \psi (\r)
:
| g \rangle,
\end{equation}
\begin{equation}
\langle x \eta | 
: {\cal H}_I^{(2)} :
| g \rangle  
\simeq
{e^2 \over 2 m c^2} A^2(\r_{x \eta})
\int_V \! \! d^3 r \ 
\langle x \eta | 
:
\hat \psi^\dagger (\r)
\hat \psi (\r)
:
| g \rangle, 
\end{equation}
and similarly for other matrix elements.
Here, $\r_{x \eta}$ denotes the center position around which 
$\varphi_{x \eta}$ is localized.
We may use, for example, $\varphi_{N \nu}$ for $\varphi_{x \eta}$, 
and evaluate the above integrals using 
(\ref{formula2}) by putting 
$Q(\r, \r') = \e \cdot \nabla \delta (\r - \r')$ 
and 
$Q(\r, \r') = \delta (\r - \r')$, respectively.
From Eqs.\ (\ref{exwf2}), (\ref{S}) and (\ref{exwf_trap}),
we find 
\begin{eqnarray}
\int_V \! \! d^3 r \ 
\langle N \nu | 
: 
\hat \psi^\dagger (\r)
\ \e \cdot \nabla \
\hat \psi (\r)
:
| g \rangle  
=
\int_V \! \! d^3 r \ 
\left[
\ \e \cdot \nabla' \
\varphi_{N \nu}^* (\r, \r')
\right]_{\r' \to \r} 
&&
\\
=
{\tilde G}_{N \nu} (\0)
{\sum_\k}' 
\tilde U_\nu^* (\k) 
(c \k | \ \e \cdot \nabla \ | v \k),
&&
\label{me1}
\end{eqnarray}
\begin{equation}
\int_V \! \! d^3 r \ 
\langle N \nu | 
: 
\hat \psi^\dagger (\r)
\hat \psi (\r)
:
| g \rangle  
=
\int_V \! \! d^3 r \ 
\varphi_{N \nu}^* (\r, \r)
=
0,
\end{equation}
where we have utilized Eqs.\ 
(\ref{norm_cell}), (\ref{norm_wannier}) and (\ref{formula0}), and 
\begin{equation}
(c \k | \ \e \cdot \nabla \ | v \k)
\equiv
{1 \over v_0} \int_{v_0} d^3 \rho \
u^*_{c \k}(\r) 
\ \e \cdot \nabla \
u_{v \k}(\r)
\end{equation}
represents contribution from each cell.
If this integral is independent of $\k$ in the range of $\k$ for which 
$\tilde U_\nu(\k)$ is not small, then we denote the integral by 
$
(c | \ \e \cdot \nabla \ | v)
$ 
and Eq.\ (\ref{me1}) is reduced to 
\begin{equation}
\int_V \! \! d^3 r \ 
\langle N \nu | 
: 
\hat \psi^\dagger (\r)
\ \e \cdot \nabla \
\hat \psi (\r)
:
| g \rangle  
=
\sqrt{V} \ {\tilde G}_{N \nu}^* (\0) U_\nu^* (\0) 
(c | \ \e \cdot \nabla \ | v)
\label{me1_red}\end{equation}
At first sight this result might look unphysical because of 
the anomalous factor $\sqrt{V}$.
However, this factor is canceled by the factor $1/\sqrt{V}$ in ${\tilde G}_{N \nu}(\0)$. 
This can be seen as follows.
Recall that we have assumed that the exciton state is 
well localized, $\ell_G < \lambda$.
For $\varphi_{N \nu}$ to 
be well localized ${\tilde G}_{N \nu}(\K)$ must be a smooth function of $\K$, and 
we can replace the summation in Eq.\ (\ref{norm_g}) with the integral as 
\begin{equation}
1 = {\sum_\K}' |{\tilde G}_{N \nu}(\K)|^2 
= {V \over (2 \pi)^3} \int d^3 K \ |{\tilde G}_{N \nu}(\K)|^2
\end{equation}
Considering also that the spatial extension of $G_{N \nu}$ is 
of the order of $\ell_G$, 
we find that the maximum value of $|{\tilde G}_{N \nu}(\0)|$ 
should behave as
\begin{equation}
\max_N |{\tilde G}_{N \nu} (\0)| \sim \sqrt{V_G \over V}
\end{equation}
where we have put, 
apart from an unimportant numerical factor of order unity, 
\begin{equation}
V_G \sim \ell_G^3
\end{equation}
Therefore, the matrix element (\ref{me1_red}) is  
independent of $V$, as it should be.
The matrix element is actually proportional to $\sqrt{V_G}$, 
which is sometimes called the ``coherence volume" 
\cite{feldmann} \cite{hanamura} because it 
gives a measure of the 
volume of the space region over which the exciton wavefunction is coherent.
An important consequence of the $V_G$ dependence of the matrix element is 
that the radiative recombination rate $\Gamma_{N \nu}^{\rm rad}$ 
of the exciton is proportional to $V_G$ in the region where
 $\ell_G < \lambda$ \cite{feldmann} \cite{hanamura}.
According to Eq.\ (\ref{ell_G}), 
this phenomenon may be observed in quantum wells and in quantum dots, 
as the dependence of $\Gamma_{N \nu}^{\rm rad}$ on temperature (on 
which $\ell_\phi$ depends) and on the 
dot size ($\simeq \ell_{loc}$), respectively. 
The size dependence of $\Gamma_{N \nu}^{\rm rad}$ of quantum dots 
was clearly observed by 
Nakamura et al.\ \cite{nakamura} and Itoh et al.\ \cite{itho},
whereas the temperature dependence of $\Gamma_{N \nu}^{\rm rad}$ of quantum 
wells reported by Feldmann et al.\ \cite{feldmann} 
could be interpreted as due to 
the temperature dependence of the exciton distribution because the experiment
was performed at relatively high temperatures \cite{akiyama}.
%
%
%

Note that the result (\ref{me1_red}) can also be obtained when 
we neglect the 
$\k$ dependence of $u_c$ and $u_v^*$ 
in Eq.\ (\ref{exwf2}) as
\begin{equation}
S_{\K \nu} (\r_e, \r_h) \simeq U_\nu (\r_e - \r_h) 
u_{c \0}(\r_e) u^*_{v \0}(\r_h)
\equiv
S_\nu^{(0)}.
\quad \mbox{(often wrong)}
\label{S_wrong}\end{equation}
In this approximation, the wavefunction $\varphi_{N \nu}$, for example, 
is decoupled into the envelope function of 
the center-of-mass motion, that of relative motion, 
and cell-periodic parts as
\begin{equation}
\varphi_{N \nu}(\r_e, \r_h)
\simeq
G_{N \nu}(\R) U_\nu (\r_e - \r_h) u_{c \0}(\r_e) u^*_{v \0}(\r_h)
\equiv
\varphi_{N \nu}^{(0)}(\r_e, \r_h).
\quad \mbox{(often wrong)}
\label{exwf_wrong}\end{equation}
This approximation has been used in much literature 
because of its simplicity.
This approximation is sometimes good for simple cases such as 
the band-to-band matrix element between direct-allowed bands
which are non-degenerate.
In studies of nonlinear optical phenomena, 
however, we must also evaluate other matrix elements. 
For such general matrix elements, 
{\it the approximate forms 
(\ref{S_wrong}) and (\ref{exwf_wrong}) 
often give wrong and unphysical results}, 
as we now explain.

As a typical example  
let us investigate the transition matrix elements between 
different exciton states. 
When the spatial extension of both the initial and final states are 
smaller than $\lambda$, 
we find from formula (\ref{formula3}) 
\begin{equation}
\langle x \eta | 
: {\cal H}_I^{(1)} :
| x \eta' \rangle  
\simeq
- {\hbar e \over i m c} \A(\r_{x \eta})
\int_V \! \! d^3 r_e  \int_V \! \! d^3 r_h \
\varphi^*_{x \eta}(\r_e, \r_h) 
\e \cdot (\nabla_e + \nabla_h)
\varphi_{x \eta'}(\r_e, \r_h).
\end{equation}
Here,  
you can arbitrarily replace $\A(\r_{x \eta})$ with $\A(\r_{x \eta'})$,  because 
$\A(\r_{x \eta}) \simeq \A(\r_{x \eta'})$ from our assumption $\lambda < \ell_G$.
The point is that 
derivatives of {\it both} the envelope functions and cell-periodic parts 
contribute to this matrix element.
For the latter, we can use the group-velocity formula;
\begin{equation}
\int_V \! \! d^3 r 
\phi_{b \k}^*(\r) {\hbar \over i m} \nabla \phi_{b \k}(\r) 
=
{1 \over \hbar} {\partial \varepsilon_b(\k) \over \partial \k},
\qquad (b = c, v)
\end{equation}
where $m$ denotes the free-electron mass.
For the parabolic dispersion 
(\ref{para_e})-(\ref{para_h}), 
this yields 
\begin{eqnarray}
(c  \k | \nabla | c \k)
&=&
i \left( {m \over m_e} -1 \right) \k,
\label{cnc}\\
(v  \k | \nabla | v \k)
&=&
i \left( -{m \over m_h} -1 \right) \k.
\label{vnv}\end{eqnarray}
It is seen that 
the derivatives of the cell-periodic parts can never be neglected 
unless $m_e = m$ and $m_h = -m$. 
For example, when $m_e \simeq m_h \simeq m$ the derivatives are
$(c  \k | \nabla | c \k) \simeq 0$ and 
$(v  \k | \nabla | v \k) \simeq -2i \k$.
Physically, these derivatives represent modification of 
velocities (whose operator is $\hbar \nabla / i m$ in the absence of  electromagnetic fields) 
due to Bragg reflections by the crystal potential.
The modification is larger for holes than for electrons because 
usually both $m_e$ and $m_h$ are positive.
Use of the simplified forms (\ref{S_wrong}) or (\ref{exwf_wrong}) 
leads to the total neglect of these derivatives 
(because they vanish at $\k=\0$), 
and often results in unphysical results.
For example, when $\varphi_{x \eta} = \varphi_{N \nu}$ we find 
\begin{eqnarray}
\langle N \nu | 
: {\cal H}_I^{(1)} :
| N' \nu' \rangle  
\simeq
- {\hbar e \over i m c} \A(\r_{x \eta})
\left[ \ 
\delta_{\nu, \nu'} 
\int_V \! \! d^3 R \ 
G_{N \nu}^*(\R) \ \e \cdot \nabla \ G_{N' \nu}(\R)
\right.
&&\nonumber\\
\left. 
+
{m \over \mu}
\int_V \! \! d^3 R \ 
G_{N \nu}^*(\R) G_{N' \nu'}(\R)
\int_V \! \! d^3 r \ 
U^*_\nu(\r) \ \e \cdot \nabla \ U_{\nu'}(\r)
\right]. && \quad
\label{exex_ok}\end{eqnarray}
Note that the first and second terms of the right hand side (rhs) 
become finite for different values of 
$N, \nu, N'$ and $\nu'$. 
Note also that the typical value of the first term is much 
smaller than that of the second term because 
$a_B^* \ll \ell_G$ (by our assumption on $\varphi_{N \nu}$) and 
$\mu < m$ (e.g., $m/\mu \sim 10$ for GaAs).
On the other hand, if we took 
$\varphi_{x \eta} = \varphi_{N \nu}^{(0)}$ 
we could obtain the first term only;
\begin{equation}
\langle N \nu^{(0)} | 
: {\cal H}_I^{(1)} :
| N' \nu'^{(0)} \rangle  
\simeq
- {\hbar e \over i m c} \A(\r_{x \eta})
\delta_{\nu, \nu'} 
\int_V \! \! d^3 R \ 
G_{N \nu}^*(\R) \ \e \cdot \nabla \ G_{N' \nu}(\R).
\label{exex_wrong}\end{equation}
It is clear that the use of the simplified form 
$\varphi_{N \nu}^{(0)}$ would lead to
wrong results for the matrix element, 
{\it both in magnitude and in the selection rules.}
Therefore,  we cannot use the simplified forms 
(\ref{S_wrong}) and (\ref{exwf_wrong}) for physical processes in which 
exciton scatterings by optical fields are involved. 
Such processes include, for example, second-order nonlinear responses and 
third-order processes in which the exciton scatterings by optical fields 
are involved as intermediate (virtual) states.

It seems worthwhile to 
point out another difficulty arising from the use of 
the simplified forms.
Equations (\ref{cnc}) and (\ref{vnv}) suggest that 
$u_{b \k}$'s play important roles in the gauge invariance of the theory.
However, the use of the simplified forms is equivalent to 
the fixing of $u_{b \k}$'s on $u_{b \0}$'s 
which are defined in some particular gauge.
As a result, {\it the gauge invariance of such a theory is violated},
\footnote{
There is another possible origin of the violation of the gauge 
invariance of the theory. It is some nonlocal interaction, 
such as the exchange interaction, 
which may appear as a result of a mean-field approximation \cite{starace}.
In our case, however, such an interaction has been neglected 
because we are considering Wannier excitons.
}
and physical predictions of the theory depend on the gauge
\cite{SGQ} \cite{PQ}.
No definite conclusions can be drawn from such a theory, of course.

Both of the above difficulties become severer 
if one proceeds from linear optical processes to nonlinear processes. 
Unfortunately, however, 
the simplified forms (\ref{S_wrong}) and (\ref{exwf_wrong}) 
have been used in much literature 
which discusses nonlinear optical processes, 
without any consideration on these difficulties. 
The most appropriate way of calculation 
is to use the correct 
wavefunctions (\ref{S}) and (\ref{exwf_trap}) together with 
the original interaction Hamiltonian (\ref{H_I}).
There are alternative ways in which one may use 
the simplified form (\ref{exwf_wrong}).
One way is to 
use of the ``length formula" for the interaction 
\cite{starace} \cite{SGQ}:
\begin{equation}
\hat {\cal H}_I^{length}
=
e \int_V \!\! d^3 r \ 
\hat \psi^\dagger (\r) \ \r \cdot {\bf E} \
\hat \psi (\r). 
\end{equation}
This expression is rather anomalous because 
the value of $\r$ extends to infinity as $V \to \infty$.
Therefore, the normally-order form $: \hat {\cal H}_I^{length} :$ should be 
used in actual calculations.
It is not difficult to show that this interaction together with 
the simplified form (\ref{exwf_wrong}) give the correct results.
The other way is to
perform calculation according to the following prescription:
In calculations of the {\it envelope functions} and matrix elements 
of {\it intraband transitions} (exciton-exciton scatterings), 
use the following Luttinger-Kohn Hamiltonian as 
an effective ``intraband Hamiltonian" {\it for 
the envelope function} $G_{N \nu} U_\nu$ \cite{SGQ};
\begin{equation}
H_{intra}^{env}
\equiv
{1 \over 2 m_e} \left( {\hbar \over i} \nabla_e + {e \over c} \A \right)^2 
+
{1 \over 2 m_h} \left( {\hbar \over i} \nabla_h - {e \over c} \A \right)^2 
- {e^2 \over \epsilon |\r_e - \r_h|}.
\label{H_intra}\end{equation}
In calculations of matrix elements 
of {\it interband transitions} (exciton creation and annihilation), 
on the other hand,
one must use the bare interaction $\hat {\cal H}_I$.
Physically, this prescription indicates that 
slowly-varying phenomena are determined by the effective masses
whereas the rapidly-varying phenomena are determined by the 
free-electron mass \cite{knox}.

\subsubsection{Quantum States in Nanostructures}
\label{QSN}

The above formulation must be modified to treat  
nanostructures, such as quantum wells (QWs) and quantum wires (QWRs),
because 
the band structures are modified by the quantum confinement, 
and also because nanostructures are usually made of semiconductors 
which have degenerate valence bands (whereas we have assumed 
for simplicity that the bands are non-degenerate).
One way to take account of these effects is to combine 
the $\k \cdot {\bf p}$ theory with the exciton theory, 
which would enable us to determine (though perturbatively) 
both the band structure and
the exciton wavefunction simultaneously.

We here describe another simpler way --- a rigid-band approximation.
This approximation may be valid (i) in the case where  
the quantum confinement is very strong 
(i.e., the confinement energy $\gg$ the exciton binding energy) 
and also (ii) in the case 
where the quantum confinement is very weak 
(i.e., the confinement energy $\ll$ the exciton binding energy) 
{\it and} 
both the $c$ and $v$ bands are non-degenerate (i.e., these bands 
are energetically separated  around $\k=\0$ from other bands, 
and the energy separation is larger than the exciton binding energy).
We can take the band structures as those of  
the nanostructure and the bulk crystal for cases (i) and (ii), 
respectively, because the band structure would be rigid against 
the e-h interaction. 
The electron and hole operators can be defined according to 
the rigid band structure.
We can then evaluate exciton states taking the e-h interaction into 
account. 
Using these exciton states, 
we can evaluate various optical properties, 
such as nonlinear susceptibilities, of the nanostructure.

In case (ii), in particular, 
the rigid bands are the bands of {\it the bulk crystal}, 
and the only important effect of the quantum confinement 
would be the confinement of the center-of-mass motion.
Hence we may employ $\varphi_{N \nu}$ of (\ref{exwf_trap}) 
as the exciton wavefunction, 
and the formulations of the previous subsections can be applied directly.
A typical example is the 1s exciton in a quantum dot whose 
size is larger then $a_B^*$ and which is 
made of non-degenerate semiconductors.
For higher energy states (2p, 3d, and so on), 
the weak-confinement 
assumption becomes worse  
because the spatial extension of the e-h relative motion becomes larger for higher energies.

In case (i), on the other hand,  
the band structure would be modified by the quantum confinement, but 
the {\it modified} band structure would {\it not} be affected 
strongly by the e-h interaction. 
In GaAs/AlGaAs nanostructures, for example, 
the two-fold degeneracy at the top of the valence bands is lifted by 
the quantum confinement, and two series of 
subbands are formed, which are called the heavy-hole subbands and 
light-hole subbands. The subbands are no longer degenerate 
at the top of the valence bands.
By the assumption that 
the intersubband energy separation is 
larger than the exciton binding energy, 
the e-h interaction would not alter the subband structure strongly.
In this case, we may take the single-body eigenstates 
of the mean-field Hamiltonian {\it of the nanostructure}
as basis functions.
For concreteness, we assume for a nanostructure 
a quantum well structure of well width $L$ and area $S$, and
$a \ll L \ll a_B^*$ ($a$ is the lattice constant).
The cases of a quantum wire and quantum dot can be formulated in 
similar manners.
Then, a single-body eigenstate  (of low energies) may be approximated by the following form;
\begin{equation}
\phi_{b n \kp}(\r)
=
\sqrt{v_0 \over S} \sum_\ell 
\exp[i \kp \cdot \R_{\ell \parallel}]
F_{b n} (Z_\ell) w_b(\r - \R_\ell; \R_\ell),
\label{QWes}\end{equation}
where $n$ denotes a quantum number, 
$\kp$ and $\R_{\ell \parallel}$ denote two-dimensional vectors 
which represent the wavenumber and the cell position, respectively, 
in the QW plane (xy plane), 
and $Z_\ell$ is the $z$ coordinate of the cell position, 
i.e., $\R_\ell = (\R_{\ell \parallel}, Z_\ell)$.
Note that $b$ here labels the bands which are modified 
(the degeneracy is lifted) by the quantum confinement, 
and $w_b(\r - \R_\ell; \R_\ell)$ denotes the Wannier function {\it of the modified band} in the well (when $\R_\ell \in$ well region) 
or barrier (when $\R_\ell \in$ barrier region).
The Wannier functions and the envelope function $F_{b n}$ 
may be obtained, e.g., by using the $\k \cdot {\bf p}$ perturbations.
In particular, 
for bands which are non-degenerate in the bulk crystal, we may approximate
$w_b$ by that of the bulk crystal, and $F_{b n}$ may be obtained by 
solving the effective-mass equation; 
\begin{equation}
\varepsilon_b \left( \kp, -i {\partial \over \partial z} ; z \right) 
F_{b n \kp}(z)
 = \varepsilon_{b n \kp} 
F_{b n \kp}(z),
\label{Fnondeg}\end{equation}
where $\varepsilon_b (\kp, k_z; z)$ is the dispersion of the bulk band
of the semiconductor which composes the well (when $z \in$ 
well region) or barrier (when $z \in$ barrier region).
For a parabolic band, in particular, 
Eq.\ (\ref{Fnondeg}) becomes
\begin{equation}
\left[
{\hbar^2 k_\parallel^2 \over 2 m_{b \parallel}(z)}
-
{\hbar^2  \over 2 m_{b z}(z)}
{\partial^2 \over \partial z^2}
+ 
\varepsilon_b(\0; z)
\right]
F_{b n \kp}(z)
= 
\varepsilon_{b n \kp} 
F_{b n \kp}(z).
\label{Fparab}\end{equation}
Here we have considered the fact 
that in general the effective mass $m_b$ 
becomes anisotropic by the band mixing.
The $\kp \to \0$ limit of this equation is the familiar form;
\begin{equation}
\left[
-
{\hbar^2  \over 2 m_{b z}(z)}
{\partial^2 \over \partial z^2}
+ 
\varepsilon_b(\0; z)
\right]
F_{b n}(z)
= 
\varepsilon_{b n} 
F_{b n}(z),
\label{Ffamiliar}\end{equation}
where
$F_{b n} \equiv F_{b n \0}$ and 
$\varepsilon_{b n} \equiv \varepsilon_{b n \0}$.
Equations (\ref{Fparab}) and (\ref{Ffamiliar}) are basically  
particle-in-a-box problems.
Note however that 
the boundary conditions on $F_{b n \k}$ or $F_{b n}$ at 
heterojunctions depend on 
the functional forms of $w_b(\r -\R_\ell; \R_\ell)$.
For example, when $w_b$ takes different values across 
a heterojunction (because the semiconductors are different on both 
sides) then $F_{b n \k}$ must be discontinuous
in order for the wavefunction to be continuous across 
the heterojunction \cite{AM}.

For a typical GaAs/AlGaAs QW, 
the conduction band is non-degenerate and  we may use 
Eq.\ (\ref{Fnondeg}), whereas the valence bands are degenerate and 
we should use another method such as the 
$\k \cdot {\bf p}$ perturbation.
However, 
Eqs.\ (\ref{Fnondeg})-(\ref{Ffamiliar}) 
have been widely used even for the valence bands.
This may be justified if we replace 
$w_b$ with that of the QW (i.e., $w_b$ which is 
evaluated with the $\k \cdot {\bf p}$ perturbation) and 
simultaneously adjust parameters 
in $\varepsilon_b (\kp, k_z; z)$ of Eq.\ (\ref{Fnondeg})
in such a way that the wavefunction agrees with the result of 
the $\k \cdot {\bf p}$ perturbation.

We can also construct the wavefunction of an exciton, 
using the Wannier functions of the QW, as
\begin{equation}
\varphi_{x \eta}(\r_e, \r_h)
=
v_0 \sum_j \sum_\ell 
F_{x \eta}(\R_j, \R_\ell)
w_c(\r_e - \R_j; \R_j) w^*_v(\r_h - \R_\ell; \R_\ell).
\end{equation}
The envelope function $F_{x \eta}(\r_e, \r_h)$ can be evaluated 
from, say, an effective-mass equation.
Using $\varphi_{x \eta}$'s thus obtained, we can evaluate various  
optical matrix elements in a similar manner to the previous subsections.
The most important point in this case is that 
the Wannier functions (and therefore the cell periodic functions) become
anisotropic when the band degeneracy is lifted by the quantum 
confinement \cite{Yama}.
For example, when the valence band degeneracy is lifted, the matrix elements
between the cell periodic functions become anisotropic, and, 
for small $k$, 
\begin{equation}
(v \k| \nabla | v \k)
=i \left( -{m \over m_h(\hat \k)} -1 \right) \k,
\end{equation}
where $\hat \k$ denotes the unit vector which is parallel to $\k$.
The interband matrix element also becomes anisotropic, which 
for small $k$ we denote as
\begin{equation}
\lim_{\k \to 0} (c \k| \e \cdot \nabla | v \k)
= (c | \e \cdot \nabla | v \hat \k).
\end{equation}
These anisotropies of the cell-periodic parts 
have been observed as anisotropies of 
various optical properties of QWs.
It is clear that as is the case of the bulk crystal (subsection \ref{OME})
optical matrix elements must be evaluated carefully.

We have calculated in 
Refs.\ \cite{S89} \cite{SOS92} \cite{OS93} \cite{SGS92}
\cite{SY94} linear as well as nonlinear optical properties 
of nanostructures using 
the above formulation. Up to now 
some of the results have been confirmed experimentally 
\cite{fujii} \cite{tai} \cite{cingolani}.

\subsubsection{Quantum Optical Phenomena in Nanostructures}

So far 
the photon field has been treated as a classical field.
In recent years, however, quantum optical phenomena in nanostructures 
have been studied by many researchers.
We here depict some examples from our work.

The first example is the quantum nondemolition
measurement of the photon number using an electron interferometer 
which is composed of double quantum wires \cite{S91} \cite{S94}.
It was shown that one can measure the photon number 
of a quantum state of photons using the electron interferometer as
a photodetector in such a way that 
the post-measurement state of the photons 
carries the same information on the photon number as the 
pre-measurement sate.
When the pre-measurement state is an eigenstate of the photon number, 
in particular, the wavefunction is unchanged by the measurement, 
whereas the observer obtains information on the photon number.
In contrast, when the observer 
uses conventional photodetector the information on the photon number is
lost by the measurement: when he uses a photodiode, for example,  
the post-measurement photon state is the zero-photon state, 
which has no information on the pre-measurement state.
In Refs.\cite{S91} \cite{S94} 
various quantities, such as the density operator of 
the post measurement state and the measurement error, 
have been calculated 
as a function of the pre-measurement state of 
the quantized photon field and 
the structural parameters of the quantum-wire interferometer.

The second example is 
the photon creation from a false vacuum of semiconductors \cite{OS95}.
In quantum field theory, particles are created when 
a parameter(s) in the Lagrangian 
(or, equivalently, in the field equations) 
have time dependence even if the initial state is the 
vacuum state (which is defined before the time variation begins)
\cite{fulling}.
Since this is a general property, 
we can expect such phenomena also in condensed matter.
For example, 
photons may be created if a material parameter(s) varies 
as a function of time \cite{Ya} \cite{Sc}.
We have pointed out that 
conventional discussions on the photon creation in condensed matter, 
which are based on phenomenological 
quantization of the macroscopic photon field, break down 
near the exciton resonance \cite{OS95}.
To treat the exciton resonance correctly, 
we have proposed the use of a two-field model, 
in which both the photon field and the exciton field is taken into account 
as microscopic degrees of freedom \cite{OS95}. 
The two-field model has allowed us to evaluate the 
creation spectra over a wide energy range.
It is also pointed out there that 
even a slight singularity in the functional form of the 
time variation of the parameters causes large deviation from 
non-singular (physical) results, whereas 
singular forms were assumed in much literature.

The third example is 
the radiative lifetime of an excited atom in an inhomogeneous and 
absorptive cavity \cite{KS96}.
We have demonstrated strong dependence of the lifetime on the 
material parameters such as the absorption coefficient and the 
cavity structure.
The forth example is the photon statistics of 
light emitting diodes (LEDs) at a low injection level \cite{FS97}.
We have derived a formula which gives the photon statistics as a function of the pump statistics, device parameters, and the detection efficiency of 
each photon mode.

These are only a very limited examples of the 
quantum optical phenomena in nanostructures.
More examples will be presented in the subsequent sections and in 
the references cited therein.

\end{document}